# Anomalous vortex Hall effect in a ferromagnet/superconductor heterostructure

Weideng Sun[1], Przemyslaw Swatek[1], Yihong Fan[1], Hwanhui Yun[2,3], Deyuan Lyu[1], K. Andre Mkhoyan[2], Jian-Ping Wang[1] and Gang Qiu[1]*

[1]Department of Electrical and Computer Engineering, University of Minnesota, Twin Cities, Minneapolis, MN, 55455, USA

[2]Department of Chemical Engineering and Materials Science, University of Minnesota, Twin Cities, Minneapolis, MN, 55455, USA

[3]Korea Research Institute of Chemical Technology, Daejeon, 34114, Korea

*Corresponding author. E-mail: gqiu@umn.edu

## Abstract

The coexistence of superconductivity and ferromagnetism is a fascinating and complex phenomenon in condensed matter physics, as these two states are typically mutually exclusive due to their competing spin configurations. However, the interplay between these two orders through the proximity effect has been a subject of intense research as it opens up possibilities for novel technological applications. Here, we report the coexistence of superconductivity and ferromagnetism in superconducting $\delta$-TaN/ferromagnetic CoFeB heterostructures grown by facing-target sputtering. Superconducting states are comprehensively investigated, with evidence of strong correlation between the superconducting and ferromagnetic order parameters. In particular, we observed an anomalous Hall signal without the presence of the magnetic field in the mixed state of the superconducting transition near the critical temperature. Systematic characterizations of the Hall resistance under varying temperatures and magnetic fields attribute this behavior to the vortex Hall effect (VHE), whereby superconducting vortices in the mixed state undergo transverse motions near the critical temperature. Unlike previously reported VHEs in conventional type-II superconductors, the anomalous VHE in TaN is induced by the stray field in the underlying CoFeB layers. The concurrency of strong spin-orbit coupling, the superconductivity in the TaN layer, and the highly spin-polarized ferromagnetic ordering in the CoFeB layer offers new insights into proximity-induced vortex dynamics and the design of novel superconducting spintronic devices.

## I. Introduction

Superconductivity and ferromagnetism are generally considered incompatible due to their competing spin configurations. Superconductivity, characterized by the absence of electrical resistance and the expulsion of magnetic fields (the Meissner effect), relies on the formation of electron Cooper pairs with opposite spins and momenta, while ferromagnetism arises from the alignment of electron spins, leading to spontaneous magnetization. The competition between superconducting and ferromagnetic states has been discovered in materials such as rare-earth compounds $ErRh_4B_4$[1] and $HoMo_6S_8$[2], where superconductivity and ferromagnetism emerge at different temperature ranges. These materials exhibit reentrant superconductivity, where superconductivity is suppressed as ferromagnetism emerges. However, experimental discoveries and theoretical advancements have unveiled materials and systems where both phenomena can coexist[3, 4], leading to fascinating and complex phenomena in condensed matter physics. For example, in uranium-based heavy fermion compounds $UTe_2$[5, 6], $UGe_2$[7], $URhGe$[8], $UCoGe$[9], and $UIr$[10], superconductivity emerges within a ferromagnetic phase. Magnetic superconducting materials like iron-based superconductors have also shown evidence of coexisting superconductivity and ferromagnetism[11-13]. In addition, superconductivity and ferromagnetism can coexist at interfaces of superconducting/ferromagnetic heterostructures due to proximity effects. The interplay between superconducting and ferromagnetic states has been a subject of intense research as it presents opportunities to create time-reversal symmetry-breaking superconductivity that is crucial for essential topological quasiparticles such as Majorana fermions[14]. Understanding the coexistence of superconductivity and ferromagnetism not only advances fundamental physics but also paves the way for innovative technologies in spintronics, quantum computing, and advanced magnetic sensors[4].

Tantalum nitride (TaN) is a metal nitride that often exhibits a higher critical temperature ($T_c$) than its elemental counterpart[15]. With strong spin-orbit coupling, TaN is a promising material for superconducting spintronics, a rapidly growing research field that has become increasingly significant for modern quantum electronics[16-18]. However, the interlayer coupling between superconducting TaN and magnetic thin films has not yet been extensively explored. Various exotic phenomena can be expected at the superconductor/ferromagnet interface, including domain wall superconductivity[19, 20], supercurrent generation via interfacial Rashba spin-orbit interaction[21], supercurrent-induced spin-orbit torque[22], and the vortex spin Hall effect[23]. One particular example is the anomalous vortex Hall effect (VHE), arising from the interplay between vortex dynamics and associated magnetic fluxes in the mixed state near $T_c$ of type-II superconductors[24-26], and it has been reported in traditional superconductors[26] and high-$T_c$ cuprates[24, 27-34]. However, to date, there is no experimental report on anomalous VHE without the presence of the external magnetic field.

Here, we report the comprehensive transport study of the superconducting TaN thin films with the ferromagnetic CoFeB layers on MgO (001) substrates. High-quality crystalline $\delta$-TaN was deposited using a unique facing-target sputtering system[18]. Temperature-, magnetic field-, and current-dependent electrical transport measurements were thoroughly performed to reveal the distinct features of the correlation between superconductivity and ferromagnetism. A suppressed lower critical field $H_{c1}$ is observed in the superconducting states as a consequence of the stray field of the magnetic CoFeB layer. The anomalous VHE without an external magnetic field in the mixed states of the TaN layers is demonstrated during the superconducting phase transition, indicating spontaneous time-reversal symmetry-breaking superconducting states through the proximity effect. The Hall signal reverses its sign as the temperature moves across $T_c$. The VHE can be ascribed to the combined influence of the normal electrons and the vortex motion induced by the Magnus force resulting from the superfluid. This research provides valuable insights into understanding the further mechanism of the vortex dynamics in type-II superconductors in superconducting/ferromagnetic heterostructures and contributes to the development of novel superconducting spintronic devices.

## II. Methods

The superconducting layer $\delta$-phase TaN (10 nm) was deposited on one-side-polished (001)-oriented MgO substrates at 400 °C under a base pressure of less than $3.5 \times 10^{-8}$ Torr. The deposition was carried out using a facing-target sputtering (FTS) system, which enables the growth of high-quality layers while minimizing radiation damage[18, 35]. Subsequently, a multilayer stack comprising CoFeB (3.5 nm)/MgO (2.5 nm)/Ta (4 nm) was grown on TaN using a six-target Shamrock magnetron sputtering system at room temperature under a base pressure of less than $5 \times 10^{-8}$ Torr. CoFeB denotes $Co_{20}Fe_{60}B_{20}$.

Cross-sectional transmission electron microscopy (TEM) samples were prepared via a focused-ion beam (FIB) lift-out method using an FEI Helios Nanolab G4 dual-beam system. Scanning transmission electron microscopy (STEM) imaging and spectroscopy were performed using an aberration-corrected FEI Titan G2 60-300 STEM equipped with a Super-X energy dispersive X-ray (EDX) detector. TEM was operated at 200 keV with a screen current of ~30 pA. The convergent semi-angle of the STEM probe was 17 mrad, and the annular dark-field detector inner angles were 21 mrad and 93 mrad for low-angle annular dark-field (LAADF) and high-angle annular dark-field (HAADF) images, respectively.

The six-terminal Hall bar devices with dimensions $10 \times 100$ μm$^2$ were patterned using MA6 contact optical lithography, followed by etching in an Intlvac Nanoquest argon ion mill etching system. After defining the channel, the contact electrodes Ti (20)/Au (100) were patterned with photolithography and deposited using an electron-beam evaporator.

The electrical transport measurements were performed by utilizing a Quantum Design Dynacool cryostat system with the temperature range of 300 to 1.7 K and the magnetic field range of ±9 T. A Keithley 6221 current source meter and SR860 lock-in amplifiers were employed for longitudinal and Hall resistance measurements with the alternating current of 10 μA and the frequency of 23.333 Hz. A Keithley 6221 current source meter and a Keithley 2182A nanovoltmeter were employed for longitudinal direct voltage measurements.

### III. Results and Discussion

Figure 1(a) presents a schematic of the MgO (001)/TaN (10 nm)/CoFeB (3.5 nm)/MgO (2.5 nm)/Ta (4 nm) multilayer stack, along with a low-magnification cross-sectional HAADF-STEM image and corresponding EDX elemental maps. The EDX maps reveal clear elemental boundaries, indicating uniform composition and thickness across each layer with minimal atomic diffusion. High-magnification HAADF-STEM images reflect detailed views of the structures and interfaces between the MgO/TaN, TaN/CoFeB, and CoFeB/MgO/Ta layers, as shown in Fig. 1(b). The image of the TaN layer shows some dark-contrast patches alongside brighter regions of lattice contrast, indicating the presence of small amorphous regions. The LAADF-STEM image of the CoFeB/MgO/Ta structure shows a partially crystalline MgO layer deposited on the amorphous CoFeB layer, which can induce some perpendicular magnetic anisotropic regions at the CoFeB/MgO interface even without the annealing process[36, 37].

To investigate the ferromagnetic properties of the stack, the anomalous Hall effect (AHE) was first measured on the fabricated Hall device at different temperatures above $T_c$ with the magnetic field perpendicular to the sample plane, as shown in Fig. 2(a). The gradual variation in the Hall resistance $R_{xy}$ with magnetic field indicates the in-plane magnetic anisotropy of the CoFeB layer with the anisotropic field of approximately 1.7 T, which is due to the large thickness of the CoFeB layer and the absence of crystallization at the CoFeB/MgO interface without an annealing process[38]. The amplitude of AHE resistance decreases as the temperature lowers, which is uncommon in typical ferromagnetic materials. This behavior has been observed in certain magnetic systems where the spin-orbit coupling plays a significant role[39, 40]. In addition to the gradual variation of Hall resistance with magnetic field, the $R_{xy}$ exhibits hysteresis within a small magnetic field range, as shown in Fig. 2(b). This behavior can be ascribed to a small component of perpendicular magnetic anisotropy (PMA) in certain regions of CoFeB, which may be due to to either the proximity effect between the strong spin-orbit coupling (SOC) TaN layer and CoFeB[18]; or some crystallized domains at the CoFeB/MgO interface (as seen Fig. 1(b)). The coercive field increases with decreasing temperature, as the thermal energy increases the random movement of magnetic domains within CoFeB at higher temperatures, making it easier for an external field to flip the magnetization[41, 42].

The superconducting properties of the TaN/CoFeB devices were investigated by measuring the longitudinal resistance $R_{xx}$ versus the out-of-plane magnetic field at temperatures ranging from 6.0 to 1.8 K with a step of 0.1 K, as illustrated in Fig. 3(a). The $R_{xx}$ value quickly deviates from zero with a small applied magnetic field (< 100 Oe, inset of Fig. 3(a)), which is different from the normal superconducting structures and related to the magnetic proximity effect[18]. The upper and lower critical fields $H_{c2}$ and $H_{c1}$ (defined as 90% and 10% of the normal resistance value, respectively) were determined from the $R_{xx}$-$H$ curves, and the $H_c$-$T$ phase diagram is shown in Fig. 3(b). The solid lines are fitted with the Bardeen-Cooper-Schrieffer (BCS) model[16, 43, 44]:

$$H_c(T) = H_c(0)[1-(\frac{T}{T_c})^2] \tag{1}$$

where $H_c(0)$ is the critical field at 0 K. It can be deduced that $H_{c2}(0)$ and $H_{c1}(0)$ are approximately 4.1 and 0.6 T, respectively. The normal, mixed, and Meissner states are also identified in Fig. 3(b) in the phase diagram. It should be noted that experimentally extracted $H_{c1}$ (black circles) significantly deviates from the fitting curve (red solid line) at temperatures right below $T_c$. This discrepancy is also reflected in the narrow resistance tails in Fig. 3(a), in contrast to that of typical type-II superconductors[45]. This can be explained by the stray field from the CoFeB layer suppressing superconducting states, especially when the external magnetic field is comparable or lower than the stray field.

To investigate the current dependence of the superconducting transition, we measured the longitudinal voltage $V_{xx}$ versus direct current (DC) under different temperatures at zero magnetic field, as shown in Fig. 3(c), and the temperature dependence of the critical current density $j_c$ is presented in Fig. 3(d). The solid line represents the temperature dependence of the de-pairing critical current fitted by the G-L theory[46]:

$$j_c(T) = j_c(0)[1-(\frac{T}{T_c})^2]^\alpha [1+(\frac{T}{T_c})^2]^\beta \tag{2}$$

where $j_c(0)$ is the de-pairing critical current density at 0 K, $\alpha = 7/6$, $\beta = 5/6$ for superconductors with defects pinning[46, 47]. Strong pinning effect is expected in the TaN/CoFeB heterostructure, as the domain walls in the magnetic layer can serve as vortex pinning centers in the adjacent superconducting layer. A $j_c(0)$ of ~ 0.28 MA/cm$^2$ can be extracted from the fitted curve. This critical current density value is noticeably lower than that reported in the literature[48], suggesting that the adjacent magnetic layer suppresses superconductivity due to the interlayer coupling between the TaN and CoFeB layers.

Figure 4(a) presents the longitudinal resistance $R_{xx}$ and Hall resistance $R_{xy}$ as a function of temperature near the superconducting transition at zero magnetic field (the inset displays a broader temperature range from 300 to 1.7 K). The $T_c$ of the TaN/CoFeB device of approximately 4.1 K (determined at half the value of the normal state resistance) can be inferred from the $R_{xx}$-$T$ curve. Upon zero-field cooling, the $R_{xx}$ gradually drops to zero, as expected for the typical superconducting transition. Meanwhile, a positive $R_{xy}$ peak emerges as the temperature approaches $T_c$ and then reverses its sign as the temperature passes across $T_c$. The $R_{xy}$ eventually vanishes as the film fully superconducts. The non-zero Hall signal at zero field cooling procedure can be attributed to the vortex motion under the influence of the underlying magnetic layer. This is a field-free version of anomalous VHE, which is significantly different from the reported VHE[24, 27-34] in that (i) it does not require an external magnetic field; (ii) it emerges both above and below $T_c$, with a reversed sign. This anomalous VHE is highly reproducible, as similar qualitative results were also observed in other TaN/CoFeB samples as shown in Supplementary Note 1. The field-free VHE can be attributed to the local stray magnetic field generated by the adjacent CoFeB layer[49], as we will elaborate on its mesoscopic origin later.

To investigate the spontaneous anomalous VHE near $T_c$, the temperature dependence of $R_{xx}$ and $R_{xy}$ were measured simultaneously under different out-of-plane magnetic fields ranging from 0 to 5.0 T, as shown in Figs. 4(b) and 4(c). The $R_{xx}$-$T$ and $R_{xy}$-$T$ curves under negative magnetic field are presented in Supplementary Note 2. Below $T_c$, $R_{xx}$ gradually deviates from zero with the magnetic field increasing, which implies that the magnetic field and the local stray field generated by the CoFeB layer can destroy the superconducting state of the adjacent TaN layer[19]. When the temperature is above 4.7 K, the superconducting state vanishes, and the normal AHE from the CoFeB dominates the Hall signal, which is consistent with the $R_{xx}$-$H$ results in Supplementary Note 3. The in-plane magnetic field-dependent electrical transport properties were also measured, and the results are shown in Supplementary Note 4. The in-plane critical field is larger than the out-of-plane case due to the dimensional effects of the superconductor and the formation energy of the vortex, which is consistent with the magnetic field angle-dependent electrical transport measurement results in Supplementary Note 5. The VHE also emerges under in-plane magnetic field as illustrated in Figs. S4(b) and S4(d). The origins of the anomalous VHEs both below and above $T_c$ are discussed as follows.

Near superconducting transition, the motion of the flux vortices at a velocity $v_L$ through a superconductor produces a transverse electric field according to Faradays' law:

$$\boldsymbol{E} = -\frac{\boldsymbol{v}_L \times \boldsymbol{H}}{c} \tag{3}$$

where $\boldsymbol{H}$ is the total magnetic field experienced by the superconducting film with combined contributions from external field and the stray field generated by the ferromagnetic layer. At zero external field, a non-zero $\boldsymbol{H}$ in the out-of-plane direction can exist, the sign of which depends on imbalanced random magnetic domain distribution in the finite-sized samples and/or the small PMA magnetic component discussed in Fig. 2(b). Therefore, this non-zero $\boldsymbol{H}$ produces a transverse electric field because of the vortex motions and drives normal state electrons to generate a VHE signal.

The sign reversal of the VHE across $T_c$ can be explained by the change of $\boldsymbol{v}_L$ directions at different temperatures. According to the Nozières-Vinen (NV) model[50], vortices moving through superfluid experiences Magnus force:

$$\boldsymbol{F} = \frac{n_s e}{c}(\boldsymbol{v}_s - \boldsymbol{v}_L) \times \overline{\boldsymbol{\phi}_0}, \tag{4}$$

where $n_s$ is the superfluid electron density, $\boldsymbol{v}_s = \boldsymbol{J}/n_s e$ is the superfluid velocity, $e$ is the electron charge, $c$ is the speed of light, and $\phi_0 = h/2e$ is the vortex flux quantum. In analogy to the case of vortex motion in superfluid $^4$He, the drag force can be expressed as[28, 51, 52]:

$$\boldsymbol{f} = -\eta \boldsymbol{v}_L - \eta' \overline{\phi_0} \times \boldsymbol{v}_L, \tag{5}$$

where $\eta$ and $\eta'$ are phenomenological drag coefficients on vortices describing interactions with thermal excitations and vortex pinning effects. For a steady state with $\boldsymbol{F} + \boldsymbol{f} = 0$, the vortex motion can be deduced[28]:

$$\boldsymbol{v}_L = \frac{n_s e}{c} D \boldsymbol{v}_s \times \overline{\phi_0} + C \boldsymbol{v}_s, \tag{6}$$

where $D$ and $C$ are coefficients:

$$D = \frac{\eta}{(n_s e/c - \eta')^2 + \eta^2}, \tag{7}$$

$$C = \frac{n_s e}{c} \frac{n_s e/c - \eta'}{(n_s e/c - \eta')^2 + \eta^2}, \tag{8}$$

At the onset of the superconducting states ($T > 4.1$ K), as the temperature decreases from 6.0 to 4.1 K, the $R_{xy}$ increases, while $R_{xx}$ decreases gradually, as shown in Fig. 4(a), which represents the onset of superconducting puddle formation amongst the normal state backgrounds. In this temperature range, the frictional drag coefficient $\eta'$ is large due to strong vortex pinning and thermal effects, thus $C$ takes a negative value. Therefore at higher tempreatures the vortices have an opposite velocity component as to the superfluid flow. At temperatures below 4.1 K, superconducting states become dominant and screen out the majority of domain walls which act as pinning centers. As a result, vortex motions experience less friction, thus $C$ changes from a negative value to a positive one. As a result, $\boldsymbol{v}_L$ has a velocity component in the same direction as $\boldsymbol{v}_s$, and the transverse electric field reverses its sign at this temperature regime and so does VHE.

### IV. Conclusions

We report the observation of concurrent ferromagnetism and superconductivity in intimately coupled TaN/CoFeB heterostructures. The impact of underlying magnetic layer on superconductivity is systematically investigated through the temperature- and magnetic field-dependent electrical transport measurements. A spontaneous anomalous VHE in the mixed state, both above and below $T_c$ is without an external magnetic field. The zero-field VHE with a sign reversal can be explained by the vortex motions using a mesoscopic model by considering the Magnus force and drag force under the impact of the stray field of the CoFeB layer. This study provides new insights into understanding the further mechanism of the vortex dynamics in type-II superconductors. Furthermore, the combination of the superconducting TaN layer with strong spin-orbit coupling and the CoFeB layer with large spin polarization can offer potential for novel superconducting spintronic device designs.

**Figures**

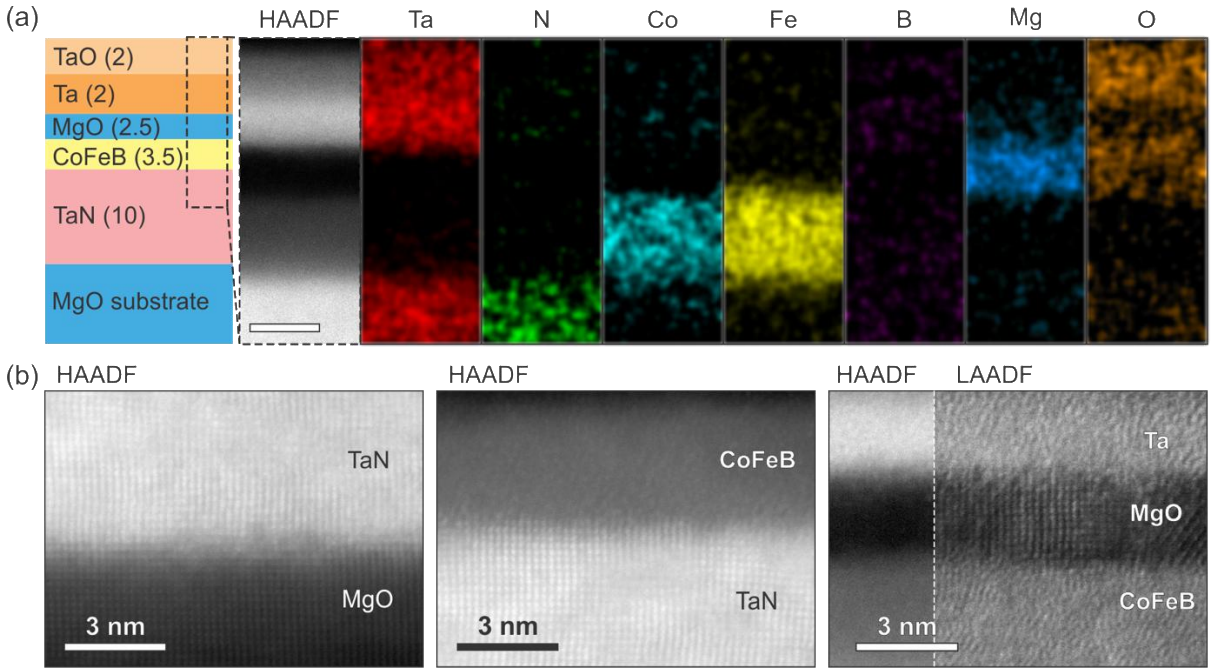

FIG. 1. Sample structure and cross-sectional STEM images. (a) Schematic of the multilayer stack and low-magnification HAADF-STEM image with EDX elemental maps. Numbers in parentheses indicate nominal thicknesses in nanometers. The scale bar is 3 nm. (b) High-magnification HAADF-STEM images showing the interfaces of MgO/TaN (left), TaN/CoFeB (middle), and CoFeB/MgO/Ta (right) layers. The LAADF-STEM image of the CoFeB/MgO/Ta structure is also presented.

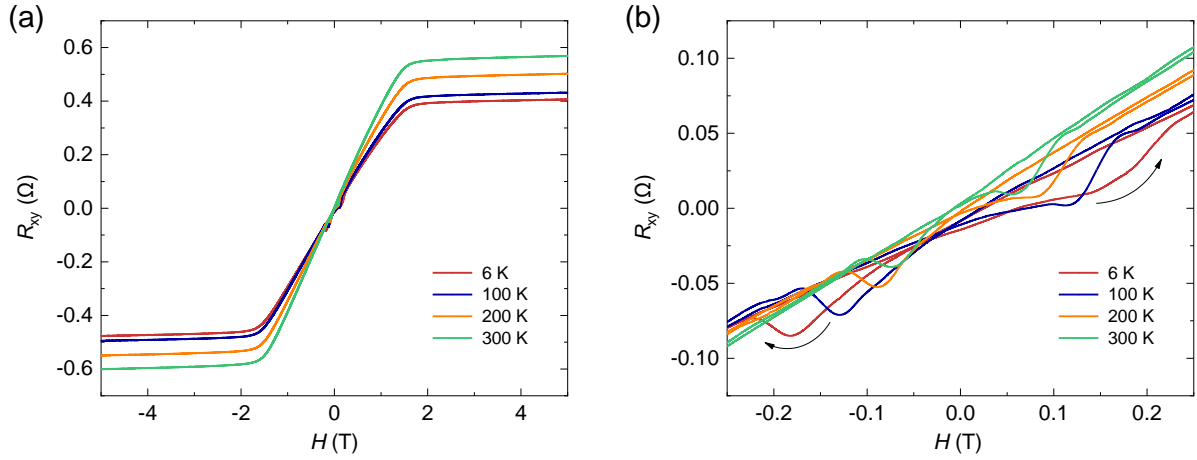

FIG. 2. The magnetic properties of the TaN/CoFeB heterostructure. (a) AHE results with the magnetic field perpendicular to the sample plane at different temperatures. (b) Zoom-in plot near zero magnetic field of (a). The black arrows indicate the sweeping direction of the magnetic field.

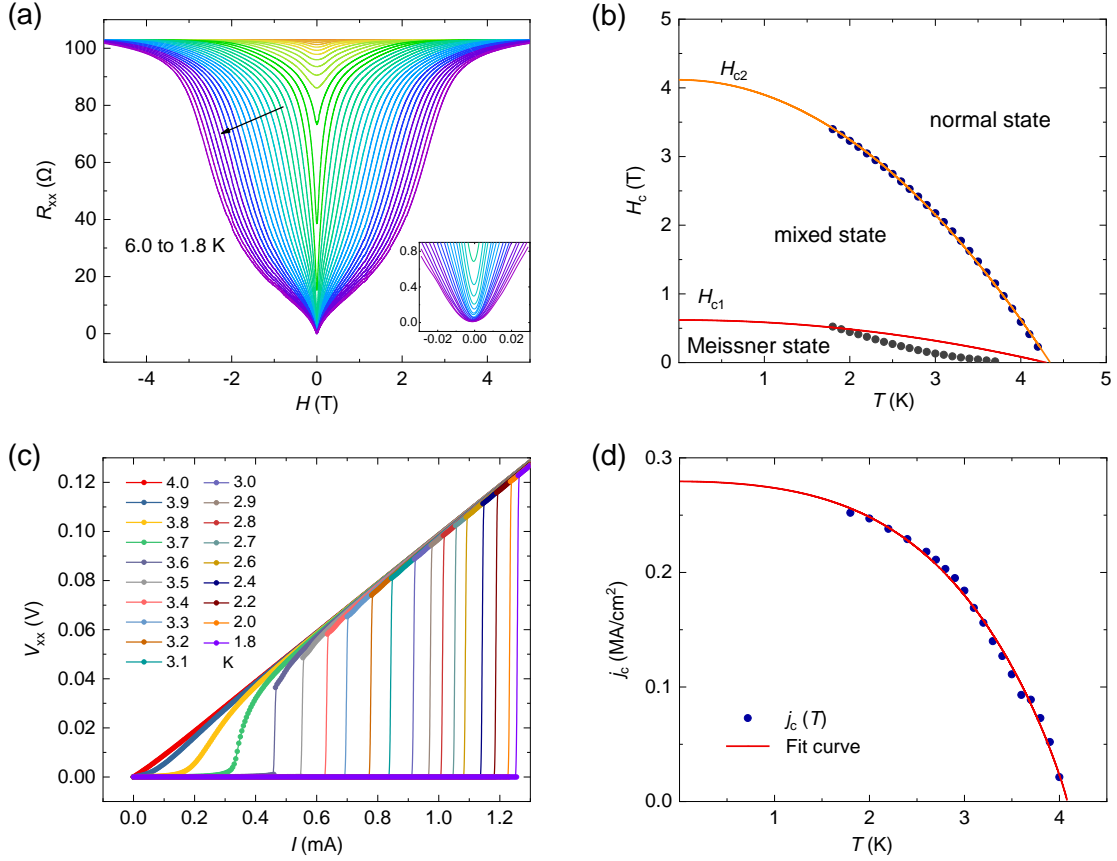

FIG. 3. Superconducting properties of the TaN/CoFeB device. (a) Longitudinal resistance $R_{xx}$ versus out-of-plane magnetic field at different temperatures. The black arrow indicates the temperature variation from 6.0 to 1.8 K with the step of 0.1 K. The inset shows the enlarged area near zero magnetic field. (b) Upper and lower critical fields $H_{c2}$ and $H_{c1}$ versus temperature phase diagrams extracted from (a). The solid lines are fitted by Eq. (1). The normal, mixed, and Meissner states are labeled. (c) Longitudinal voltage $V_{xx}$ versus direct current under different temperatures at zero magnetic field. (d) Critical current density extracted from (b) as a function of temperature. The solid lines are fitted by Eq. (2).

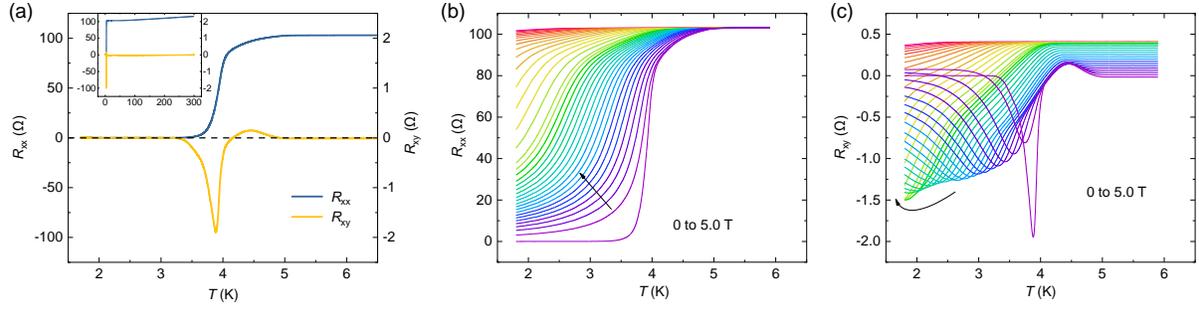

FIG. 4. Temperature- and magnetic field-dependent electrical transport measurement results of the TaN/CoFeB device. (a) Longitudinal resistance $R_{xx}$ and Hall resistance $R_{xy}$ variations with temperature. The dashed line indicates zero resistance. The inset shows the large temperature range. (b), (c) $R_{xx}$ and $R_{xy}$ versus temperature under different out-of-plane magnetic fields. The black arrows indicate the magnetic field variation from 0 to 5.0 T. From 0 to 2.0 T, the magnetic field step is 0.1 T; from 2.2 to 5.0 T, the magnetic field step is 0.2 T.

**Acknowledges**

G. Qiu acknowledges the Department of Electrical and Computer Engineering and the College of Science and Engineering at the University of Minnesota, Twin Cities for the start-up funding support.

**Author Declarations**

  **Conflict of Interest:** The authors have no conflicts to disclose.

  **Author Contributions:** G.Q., J.-P.W., and W.S. conceptualized the work. G.Q. and W.S. conceived the experiment plan. J.-P.W., P.W.S., Y.F., and D.L. grew all samples. W.S. patterned the devices and carried out electrical transport measurements. H.Y. and K.A.M. conducted STEM measurements. W.S. and G.Q. wrote and discussed the original draft. W.S., G.Q., and J.-P.W. had discussions and outlines. All authors participated in the review and editing of the manuscript.

**Data Availability**

The data that support the findings of this study are available from the corresponding author upon reasonable request.